\documentclass[conference,10pt,letterpaper]{IEEEtran}
\usepackage[utf8]{inputenc}
\usepackage[T1]{fontenc}
\usepackage[protrusion=true,expansion=false]{microtype}
\usepackage{graphicx,booktabs,tabularx,array,amsmath,amssymb,xcolor}
\usepackage{tikz}
\usetikzlibrary{arrows.meta,positioning,calc,fit,plotmarks}
\usepackage{cite}
\usepackage[hidelinks]{hyperref}
\usepackage{url}
\definecolor{mqnavy}{HTML}{173F6D}
\definecolor{mqblue}{HTML}{2C6EAA}
\definecolor{mqteal}{HTML}{287D78}
\definecolor{mqgreen}{HTML}{2E7D32}
\definecolor{mqorange}{HTML}{C46A14}
\definecolor{mqred}{HTML}{B23A2B}
\definecolor{mqgray}{HTML}{66727D}
\definecolor{mqgrid}{HTML}{D9E0E6}
\definecolor{mqlightblue}{HTML}{EAF1F8}
\definecolor{mqpaleblue}{HTML}{EDF5FB}
\definecolor{mqpaleteal}{HTML}{ECF7F6}
\definecolor{mqpalegreen}{HTML}{EFF8F0}
\definecolor{mqpaleorange}{HTML}{FFF6EA}
\hypersetup{
 pdftitle={HPC-MQBench: Qualification-First Benchmarking on Slurm with a Single-Broker Kafka Evaluation},
 pdfauthor={Sepehr Mahmoodian and Julian Kunkel},
 pdfsubject={HPC-MQBench arXiv preprint},
 pdfkeywords={Slurm, messaging, Kafka, performance measurement, qualification}}
\newcolumntype{Y}{>{\raggedright\arraybackslash}X}
\newcolumntype{L}[1]{>{\raggedright\arraybackslash}p{#1}}

\title{HPC-MQBench: Qualification-First Benchmarking\\
on Slurm with a Single-Broker Kafka Evaluation}
\author{\IEEEauthorblockN{Sepehr Mahmoodian\textsuperscript{1,2}\quad
Julian Kunkel\textsuperscript{1,2}}
\IEEEauthorblockA{\textsuperscript{1}\,University of G\"ottingen, G\"ottingen, Germany\\
\textsuperscript{2}\,Gesellschaft f\"ur wissenschaftliche Datenverarbeitung mbH G\"ottingen (GWDG), G\"ottingen, Germany\\
\href{mailto:s.mahmoodian@gwdg.de}{s.mahmoodian@gwdg.de}\quad
\href{mailto:julian.kunkel@gwdg.de}{julian.kunkel@gwdg.de}}}

\begin{document}
\maketitle

\begin{abstract}
Messaging experiments on high-performance computing clusters must coordinate services, clients, and
measurement within scheduler allocations. We present HPC-MQBench, a Slurm-orchestrated
benchmark that links resumable experiment control to record checks, delivery
accounting, and qualification before rate ranking. Its Kafka evaluation placed
producers/controller, one broker, consumers, and monitoring on four nodes. With
memory-backed logs, 120 workload configurations yielded 99 qualified observations,
14 outside the producer-delivery policy, and seven with invalid evidence.
Two validation stages each repeated ten workloads in five blocks. In the final
stage, the selected workload qualified in all five observations, with medians
of 2,927 mebibytes per second balanced endpoint rate, 1.24 percent pending
deliveries, and 2.56 seconds for the 99th-percentile latency.
It qualified in only two observations in the earlier stage, where qualification
improved after an allocation boundary without establishing a causal allocation
effect. Selection is conditional on the stage and delivery policy. Resource measurements did not isolate a
unique bottleneck. The contribution is a framework for distributed experiment control and
auditable configuration selection at a fixed broker count; multi-broker
support and scaling experiments remain future work.
\end{abstract}


\section{Introduction}
Scientific instruments and high-performance computing (HPC) telemetry services use messaging to decouple
data producers from downstream analysis~\cite{mukai2018ess,adamson2023stream}.
Selecting a messaging configuration requires more than a high producer rate:
records can still await delivery, and latency can grow while throughput appears
favourable. Tail latency matters even when typical response times are
low~\cite{dean2013tail}. On a scheduled cluster, services and clients must also
start on the right nodes, respect allocation limits, and retain enough state
to recover an interrupted experiment.

We connect these concerns in HPC-MQBench, which records placement and settings,
measures both endpoints, checks records and delivery accounting, and applies a
declared producer-delivery policy before ranking rates. A completed Slurm job
does not by itself establish an eligible measurement. An eligible case is
qualified if it meets the policy and overdriven otherwise; failed measurement
checks make a case ineligible. This separation makes selection inspectable.

The contributions are: (1) a Slurm-managed experiment lifecycle with checked
inputs, outputs, and resumable state; (2) an explicit contract separating
endpoint rates, sampled latency, record checks, and pending deliveries; and
(3) a single-broker Kafka evaluation covering workload screening, repeated
qualification, and resource evidence. The evaluation distributes experiment
roles across four nodes and uses parallel MPI clients. It examines workload
and service settings at a fixed broker count. Neither more client ranks nor
concurrent experiment batches demonstrate broker scaling, and the current
Kafka adapter supports only one broker.

\section{Related Work}
Kafka combines partitioned logs, batching, and pull-based
consumption~\cite{kreps2011kafka}. Studies of ingestion limits, client settings,
deployment choices, and delivery guarantees show why performance depends on
the workload and configuration~\cite{hesse2020kafka,fu2021fair,dobbelaere2017kafka}.
OpenMessaging Benchmark provides messaging drivers and distributed workload
workers; SPECjms2007 prescribes validity and disclosure
rules~\cite{openmessagingbenchmark,specjms2007}. Stream-processing benchmarks
address continuous queries, workload design, latency, and result
validation~\cite{arasu2004linearroad,karimov2018streambench,hesse2021espbench}.
Our focus is messaging delivery and experiment control within Slurm allocations.
In particular, the producer-callback pending ratio used here differs from
OpenMessaging's sent-minus-received backlog adjusted for subscriptions.

Pilot-Streaming manages streaming frameworks on HPC resources, Mofka targets
event streaming on supercomputers, and ADIOS~2 supports scientific data
movement~\cite{chantzialexiou2018pilotstreaming,dorier2025mofka,godoy2020adios2}.
ReFrame and Popper address scheduled tests and executable
provenance~\cite{karakasis2020reframe,jimenez2017popper}. Work on measurement
bias and statistical benchmarking motivates retaining experimental context
and reporting variation~\cite{mytkowicz2009wrongdata,kalibera2013rigorous,hoefler2015scientific}.
Validity rules and workload generators are established ideas. Our contribution
is their enforced connection to messaging-specific evidence and workflow
state: checkpoint reuse requires saved inputs and outputs to pass checks,
measurement validity determines eligibility, and qualification precedes rate
ranking. We do not measure an advantage over other benchmark harnesses.

\section{Benchmark Design}
\label{sec:design}
\subsection{Deployment and experiment lifecycle}
\begin{figure*}[!t]
\centering
\begin{tikzpicture}[
 >=Latex, font=\rmfamily\footnotesize,
 box/.style={draw,rounded corners=2pt,text width=4.2cm,minimum height=1.45cm,align=center,inner sep=5pt},
 band/.style={draw=mqnavy,fill=mqlightblue,rounded corners=2pt,text width=14.45cm,align=center,inner sep=6pt}
]
\node[band] (control) {\textbf{Slurm workflow}\quad allocation, role placement, case manifests, checkpoints, and resume};
\node[box,draw=mqorange,fill=mqpaleorange,below=8mm of control] (broker)
 {\textbf{Kafka broker node}\\broker process and topics\\RAM-backed logs; JMX};
\node[box,draw=mqblue,fill=mqpaleblue,left=8mm of broker] (producer)
 {\textbf{Producer/controller node}\\MPI control and producers\\one client per worker rank};
\node[box,draw=mqgreen,fill=mqpalegreen,right=8mm of broker] (consumer)
 {\textbf{Consumer node}\\MPI consumers\\latency and record checks};
\node[band,draw=mqteal,fill=mqpaleteal,below=9mm of broker] (monitor)
 {\textbf{Monitoring/results node}\quad Prometheus, host/process metrics, per-case JSON, campaign CSV};
\draw[->,thick,mqblue] (producer.east) -- (broker.west);
\draw[->,thick,mqblue] (broker.east) -- (consumer.west);
\foreach \n in {producer,broker,consumer} {
 \draw[->,dashed,mqgray] (control.south -| \n.north) -- (\n.north);
 \draw[->,dashed,mqteal] (\n.south) -- (monitor.north -| \n.south);
}
\end{tikzpicture}
\caption{Evaluated four-node deployment with one Kafka broker. Solid arrows carry
records; dashed lines show control and monitoring. Exporters run beside the
resources they observe. The diagram shows distributed experiment roles at a
fixed broker count.}
\label{fig:system-architecture}
\end{figure*}
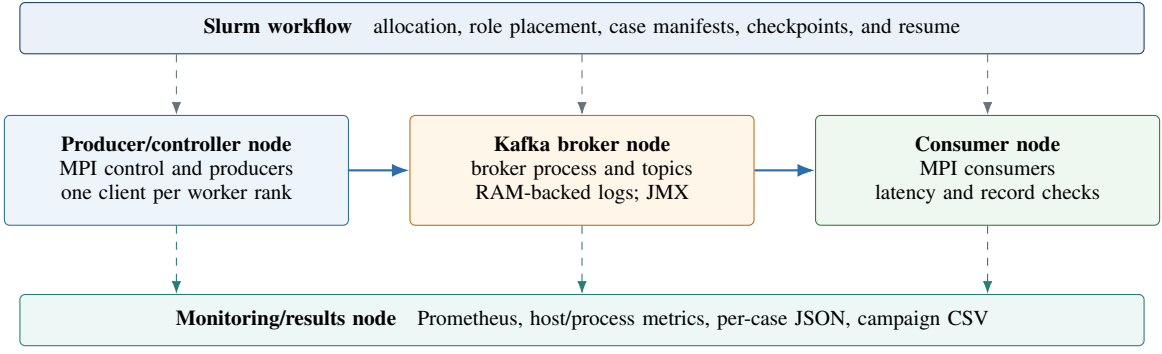

The evaluated deployment uses four exclusive nodes for producer/controller,
Kafka broker, consumer, and monitoring/results (Fig.~\ref{fig:system-architecture}).
MPI rank~0 coordinates workers through \texttt{mpi4py}~\cite{dalcin2021mpi4py};
each active producer or consumer rank owns a Kafka client. Common code controls
timing, record identity, result schemas, and qualification. The Kafka adapter
supplies \texttt{confluent-kafka}/librdkafka clients, manages broker startup and
shutdown, creates topics, and records effective settings.

The workflow validates manifests, requests allocations, places roles with Slurm
job steps, and runs cases sequentially within each
batch~\cite{jette2003slurm}. Independent screening batches can run concurrently.
Kafka restarts and its configured RAM-backed log storage is cleared before
each case. Stages record settings, job IDs, timestamps, paths, and checksums.
Resume reuses a completed stage only when its inputs and output files pass
checks; missing outputs or changed inputs stop reuse. Scheduler failures remain
workflow failures, distinct from completed cases with invalid measurements.

Prometheus collects Kafka/JVM metrics through JMX Exporter, with host and Kafka
exporters supplying additional metrics where
available~\cite{prometheus,kafkamonitoring}. A separate Python monitor reads
procfs for broker CPU/RSS, sysfs for interface bytes, and \texttt{statvfs} for
tmpfs occupancy. Process sampling, scrapes, and range-query steps are configured
at one second. Producer callback counters supply the paper's pending-delivery
ratio; it is not an exporter measurement of consumer-group lag.

\subsection{Analysis and generated outputs}
Each case writes \texttt{final\_report.json}, available monitoring data, and an
event timeline. Stage analysis checks evidence, calculates rates, classifies
cases, and produces screening, validation, and ranked-summary tables in CSV
and JSON. Campaign reports record the selected workload and broker-profile
decisions. Configuration manifests, checkpoints, job IDs, report paths, and
hashes connect decisions to their inputs. Resource summaries and diagnostic
flags aid inspection but do not automatically identify a physical bottleneck.

Latency is exported as histograms and summary statistics, including percentiles;
individual-message latency records are not exported. The full workflow selects
machine report mode. Standalone presentation modes can also create Markdown,
HTML, and plots. The manuscript's parameter contrasts, policy and endpoint-rate sensitivity,
and block comparisons are downstream analyses of retained campaign data.

\section{Measurement and Qualification}
\label{sec:measurement}
\subsection{Timing, records, and latency}
Each case uses a 15 s warm-up and 30 s measurement interval. Producers then stop
submitting, count unresolved callbacks, and flush with a 30 s timeout. Delivery
callbacks report success or failure to the application~\cite{librdkafkaintroduction}.
Consumers poll during flush and drain for up to 60 s afterwards to account for
confirmed measurement records. Warm-up records are excluded. These are
short-window measurements; no duration experiment establishes steady state.

A 32-byte envelope contains producer identity, sequence, flags, and a checksum
of envelope metadata. Consumers validate envelopes, check repeated or regressing
topic-partition offsets within each consumer, and compare total consumption
through drain with producer-confirmed deliveries. These checks detect invalid
records and count discrepancies, but do not verify full payloads or globally
reconcile record identities. Offsetting losses and duplicates can escape
aggregate counts; these checks do not prove exactly-once delivery.

Each producer timestamps measurement records whose sequence number is divisible
by ten, after payload generation and before envelope encoding and submission. Reception is timestamped
after polling and envelope/offset checks. Latency includes client buffering and
consumer polling, covering sampled deliveries through flush and drain; it
excludes payload generation and rate-scheduler waiting. The consumer rate,
however, counts arrivals only by its measurement deadline.

Before and after each case, 100 MPI ping-pong exchanges per worker estimate
offset from rank~0. The minimum-round-trip sample supplies the midpoint offset
and half-round-trip uncertainty estimate. The initial offset is used throughout.
Validity requires every active rank's estimated uncertainty and offset change
to be at most 250~$\mu$s, no negative latency, and a nonzero sample count equal
to producer-confirmed sampled deliveries. These gates do not guarantee
250~$\mu$s end-to-end accuracy.

Consumer histogram counts are summed across ranks. For percentile $p$ and $N$
pooled samples, the estimate is the first bucket upper bound reaching
$\lceil pN/100\rceil$ cumulative samples, without interpolation. Buckets become
coarser at larger latencies; a percentile in the above-60 s overflow uses the
observed maximum. Validation summaries take medians of per-run percentiles,
not a percentile of samples pooled across runs.

Fixed-size synthetic payloads derive from JSON templates with round-robin
source IDs; the envelope replaces the first 32 bytes, preserving record size.
Each producer rank has 50,000 logical IDs sharing one client and rate controller.
They are not separate device connections. The sequence includes warm-up records;
sampling therefore selects a fixed subset of 5,000 IDs per producer. Source IDs
are record keys. Samples are neither randomized nor stratified by partition;
retained summaries do not establish partition coverage or representativeness.
Consumer ranks share a case-specific group and divide partitions. Maximum offered load is limited by payload construction,
client calls, and callback polling, rather than an independent arrival stream.

\subsection{Endpoint rates and delivery policy}
\begin{figure}[t]
\centering
\begin{tikzpicture}[>=Latex,font=\rmfamily\footnotesize,x=1cm,y=1cm]
% Independent local axes; positions are schematic, not globally aligned times.
\node[anchor=west,text=mqblue] at (0,2.65) {Producer: measurement sends, then local flush};
\draw[->,thick,mqblue] (0.3,1.7) -- (7.7,1.7);
\foreach \x in {0.5,3.5,5.8} \draw[mqblue] (\x,1.61) -- (\x,1.79);
\node[below,align=center] at (0.7,1.58) {start};
\node[below,align=center] at (3.5,1.58) {stop sends\\count pending};
\node[below,align=center] at (5.8,1.58) {flush ends};
\draw[<->,mqblue] (0.5,1.95) -- node[above] {local producer duration} (5.8,1.95);
\node[anchor=west,text=mqteal] at (0,0.62) {Consumer: measurement, flush waiting, then drain};
\draw[->,thick,mqteal] (0.3,-0.15) -- (7.7,-0.15);
\foreach \x in {0.7,3.1,5.8,7.5} \draw[mqteal] (\x,-0.24) -- (\x,-0.06);
\node[below] at (0.7,-0.28) {start};
\node[below,align=center] at (3.1,-0.28) {deadline\\$B_c$ stops};
\node[below,align=center] at (5.8,-0.28) {all producers flushed\\drain begins};
\node[below] at (7.5,-0.28) {end};
\draw[<->,mqteal] (0.7,0.08) -- node[above] {local duration: 30 s} (3.1,0.08);
\node[above] at (4.45,-0.10) {keep polling};
\draw[->,dashed,mqgray] (6.9,1.64) -- (6.9,0.43) -- (5.8,0.43) -- (5.8,-0.02);
\end{tikzpicture}
\caption{Accounting after warm-up, schematically and not to scale. Each axis
has worker-local boundaries; aligned positions do not imply a shared window.
$B_p$ counts bytes from measurement sends confirmed through local flush;
$B_c$ counts bytes from measurement-tagged arrivals by the consumer deadline. Latency continues through
flush waiting and drain. The dashed arrow denotes collective coordination:
drain starts only after all producers finish flushing.}
\label{fig:accounting-windows}
\end{figure}
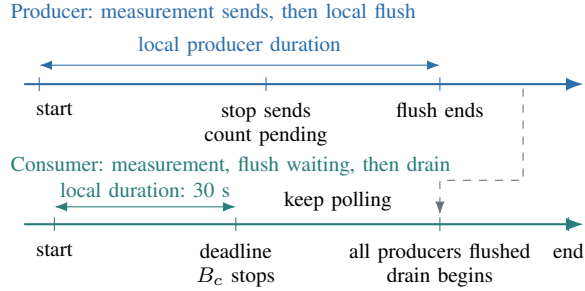

Let $B_p$ count bytes from measurement-period sends whose delivery is confirmed,
including during flush, and $B_c$ count measurement-tagged bytes received by the
consumer deadline. Both counters sum ranks. Producer duration $T_p$ is the
maximum rank duration from measurement start through flush completion; consumer
duration $T_c$ is the maximum recorded measurement duration, excluding drain.
Fig.~\ref{fig:accounting-windows} separates these boundaries. Rates in MiB/s are
\begin{equation}
\begin{aligned}
R_p&=\frac{B_p}{T_p2^{20}},\qquad R_c=\frac{B_c}{T_c2^{20}},\\
R_{\mathrm{bal}}&=\min(R_p,R_c).
\end{aligned}
\label{eq:balanced-throughput}
\end{equation}
For fixed record size $S$ bytes, $R_{\mathrm{records}}=R_{\mathrm{bal}}2^{20}/S$.
Byte rates include the envelope but exclude keys and protocol overhead.
$R_{\mathrm{bal}}$ is a \emph{balanced endpoint rate}: its minimum prevents a
faster endpoint alone from increasing the score. Because the windows differ,
it is not synchronized end-to-end throughput: flush duration affects the score
as well as data movement. The endpoint-rate ratio is not consumer lag. Retained
summaries cannot reconstruct a common-window producer rate;
Section~\ref{sec:results} compares the available endpoint scores.

Pending callbacks at flush start and failed sends are expressed as numeric
percentage values, so 5 denotes 5\%:
\begin{equation}
\begin{aligned}
D_{\mathrm{pending}}&=100\frac{N_{\mathrm{pending}}}{N_{\mathrm{attempts}}},\\
F_{\mathrm{send}}&=100\frac{N_{\mathrm{failed}}}{N_{\mathrm{attempts}}}.
\end{aligned}
\label{eq:pending-delivery}
\end{equation}
$N_{\mathrm{attempts}}$ includes all measurement-period send attempts;
$N_{\mathrm{failed}}$ includes immediate submission errors and failed callbacks.
Counts sum producer ranks. For each rank, pending callbacks are enqueued
measurement records minus observed measurement delivery callbacks, clamped at zero.
The legacy CSV field \texttt{pending\_backlog\_percent} therefore measures
producer delivery pressure, not broker queue depth or consumer-offset lag.
Measuring at send-loop stop captures unresolved work before flush can hide it;
it neither measures peak backlog nor replaces a backlog time series.

Execution, contract, backend-health, latency, or record-check failures make a
case \emph{ineligible}. An eligible case is \emph{qualified} when
\begin{equation}
D_{\mathrm{pending}}\leq5,\quad T_{\mathrm{flush}}\leq10\ \mathrm{s},\quad
F_{\mathrm{send}}\leq0.1,
\label{eq:qualification-policy}
\end{equation}
where $T_{\mathrm{flush}}$ is the maximum rank duration from send-loop end to
flush completion. An eligible case exceeding a limit is \emph{overdriven}.
These illustrative limits have no recorded application requirement and impose
no latency target. Selection is conditional on this policy; it is neither a
Kafka capacity claim nor an application service-level objective. We compare
three pending thresholds rather than asserting a universally preferable policy.
Accounting uses confirmed deliveries; permitted failed sends are not missing
delivered records. The inspected send loop advances after an immediate failure
without application-level resubmission. Requiring every attempt to succeed
would need a zero-failure policy.

Repeated validation runs each shortlisted workload once in each of five
randomized blocks. Ranking prioritizes qualified-observation count, then median
balanced rate, followed by rate IQR, pending deliveries, flush, failures, and ID
as tie-breaks. Summaries include all eligible observations, including overdriven
ones. For five observations, IQR under the inclusive convention is the fourth
minus the second ordered value. A case execution is the observation unit; allocation is
a clustering factor. Counts and medians describe these observations, not
independent estimates of qualification probability or tests of superiority.

\section{Experimental Setup}
\label{sec:setup}
The campaign used exclusive four-node allocations on GWDG's \texttt{standard96s}
partition, with two-hour job limits. Table~\ref{tab:configuration} separates
published hardware specifications from retained settings. Active SMT, binding,
CPU clocks, and a complete node inventory were not retained. Recorded runtimes
are Slurm 26.05.1, Python 3.11.9, Open MPI 5.0.6, and OpenJDK 17.0.11; exact
Kafka/client versions are unavailable. The deployment uses one KRaft broker,
replication factor one, \texttt{acks=1}, no compression, and RAM-backed logs.

\begin{table}[t]
\caption{System context and baseline settings.}
\label{tab:configuration}
\centering\footnotesize
\begin{tabularx}{\columnwidth}{@{}L{0.28\columnwidth}Y@{}}
\toprule
Property & Specification or setting \\
\midrule
Published platform & GWDG Emmy Phase 3; Rocky Linux 8~\cite{gwdgcpupartitions} \\
CPU per node & Two Xeon Platinum 8468; 96 physical cores; 2.10 GHz base, 3.80 GHz maximum turbo~\cite{gwdgcpupartitions,intel8468} \\
Memory / fabric & 514,000 MB RAM; 100 Gbit/s Omni-Path~\cite{gwdgcpupartitions} \\
Allocation / logs & Four exclusive role nodes; \texttt{/dev/shm} logs \\
Broker B0 & 8 network / 16 I/O threads; 8 GiB heap; G1 GC; 1 MiB socket buffers; 1 GiB log segments \\
Workload baseline & 40 producer / 40 consumer ranks; 120 partitions; 4 KiB records \\
Producer per client & 1 MiB batch; 20 ms linger; queue limits 1,000,000 messages and 1,024 MiB \\
Consumer per client & Fetch minimum 1 MiB, maximum 8 MiB, wait 50 ms \\
\bottomrule
\end{tabularx}
\par\smallskip\raggedright
Hardware/OS entries are published specifications, not measured campaign
frequencies, memory availability, or negotiated link rates. Other entries are
retained settings. The monitored interface name \texttt{ib0} does not identify
the fabric. MiB and KiB are $2^{20}$ and $2^{10}$ bytes.
\end{table}

\textbf{V1} denotes screening and initial validation under baseline profile B0;
\textbf{V2} denotes the subsequent instrumentation check, calibration, broker
profile comparison, and final validation. These are campaign labels, not Kafka
versions. Screening comprises 120 unique configurations: a baseline, 43
one-factor contrasts, nine additional producer/consumer rank pairs, and 67
mixed configurations sampled with seed 42 after deduplication. Four jobs each
run up to 30 cases. Ranks vary from 16 to 64 per endpoint, partitions from 60
to 240, payloads from 1 to 16 KiB, batches from 256 KiB to 4 MiB, and linger
from 0 to 80 ms; queue and fetch settings also vary. This is not a full factorial
design. One-factor cases hold the other ten settings at baseline.

The ten-workload shortlist combines the top three qualified cases, the nearest
pending-limit overdriven case passing the other limits, the highest-rate
overdriven case, eligible byte- and record-rate leaders, the baseline, and the
strongest one-factor case. Deduplication and qualification-ranked fills yield
ten workloads. Both stages use this shortlist. The inspected generators each
initialize \texttt{random.Random(20260728)} once per stage, then shuffle a fresh
shortlist copy per block, advancing the generator across blocks. Separate
initialization with the same seed reproduces the identical V1/V2 orders in the
retained manifests; this is a replay, not a recovered campaign seed log. Each
stage spans two allocations: blocks 1--3 in one and blocks 4--5 in the other.
The artifact maps V1 to jobs 15303484/15303485; V2 retains batch membership
but lacks an exact batch-to-job mapping.

The original \texttt{cfg\_103} run and both executed reruns were ineligible.
The selected 120-row screen substitutes its final rerun for the original row
and retains 119 other original rows. Seven cases remain ineligible under
unchanged rules. Workflow history records the repairs
and input canonicalization without changing raw results. The 291 historical
executions comprise 120 original screening runs, two repairs, 100 validation
runs, and 69 auxiliary runs: six instrumentation, three calibration, 30 profile
screening, and 30 additional confirmation runs. These are campaign-specific
counts, not fixed requirements of every benchmark invocation.

Before Kafka starts, \texttt{iperf3} probes TCP throughput once per allocation
in each direction, producer to broker and broker to consumer~\cite{iperf3}.
The inspected collector defaults to ten seconds and four parallel streams;
original commands and raw JSON are missing. These are not RDMA measurements.
Inventory code also queries OS/device information and runs a local STREAM-style
memory probe, but absent raw inventory prevents reporting measured hardware or
memory-bandwidth results from those probes.

\section{Kafka Evaluation}
\label{sec:results}
\subsection{Screening: rates and delivery pressure}
\begin{figure*}[!t]
\centering
\includegraphics[width=\linewidth]{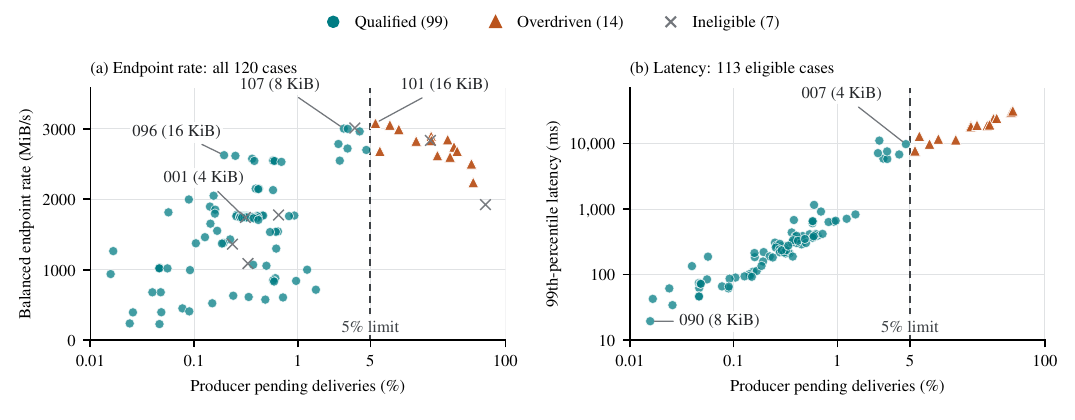}
\caption{Kafka screening results. (a) All 120 observations: balanced endpoint rate against
producer pending-delivery ratio. (b) The 113 eligible observations: p99 latency
against the same ratio; seven invalid latency observations are omitted. The dashed
line marks the 5\% policy limit. Selected labels give configuration ID and
payload size. Every point is one run; both horizontal axes and the latency axis
are logarithmic. All pending ratios are positive (minimum 0.0157\%); no zeros
are omitted or clipped.}
\label{fig:kafka-operating-limits}
\end{figure*}

Of 120 selected screening observations, 113 are eligible and 99 qualify.
Seven fail latency-validity checks; one also reports 83,350 unexplained surplus
records. Fig.~\ref{fig:kafka-operating-limits} relates rate and p99 to pending
deliveries. The baseline achieves 1,767.1 MiB/s, 269 ms p99, and 0.296\% pending.
The largest observed rate, 3,080.0 MiB/s from \texttt{cfg\_101}, exceeds the
pending limit at 5.599\%. The highest qualified byte-rate case,
\texttt{cfg\_107}, reaches 3,001.5 MiB/s with 2.776\% pending and 5,890 ms p99.

Different objectives favour different workloads. With 1 KiB records,
\texttt{cfg\_094} gives the highest qualified balanced record rate:
1,023,762 records/s, 999.8 MiB/s, and 712 ms p99. With 8 KiB records,
\texttt{cfg\_090} gives the lowest qualified p99, 19.5 ms, at 938.3 MiB/s.
The 16 KiB \texttt{cfg\_096} reaches 2,624.8 MiB/s and 221 ms p99 but
167,984 records/s. A byte-rate ranking obscures these trade-offs.

All 14 eligible overdriven cases exceed 5\% pending deliveries; six also exceed
10 s flush and two exceed 0.1\% failed sends. The overlapping groups are not
additional rejections. Qualified cases have median pending deliveries of
0.320\% and p99 of 259 ms; overdriven medians are 20.755\% and 19,000 ms.
Among eligible 16 KiB cases with at least 104 total client ranks, none of five
qualify, despite a median 2,743.0 MiB/s: median pending is 32.00\% and p99
24.2 s. These are different configurations, not repetitions of one workload.
All 99 qualified cases have zero failed sends. Passing record checks after
drain does not show that consumers kept pace during measurement.

Changing only the pending threshold from 2\% to 5\% to 10\% yields 92, 99,
and 103 qualified cases, with leaders \texttt{cfg\_096}, \texttt{cfg\_107},
and \texttt{cfg\_101}, respectively. Tightening failed sends to zero changes
nothing. These are sensitivity checks on fixed observations, not new operating
conditions. Qualified p99 reaches 11.1 s because the policy has no latency limit.

Substituting either retained endpoint rate preserves the qualified screening
leader: \texttt{cfg\_107} has $R_p=3{,}001.5$ and $R_c=3{,}118.2$ MiB/s.
The windows remain different; this comparison does not recover synchronized
throughput.

With 64 rather than 40 producer ranks, \texttt{cfg\_007} reaches 2,698.7 MiB/s
in screening (+52.7\% versus baseline), but p99 grows to 9,810 ms and pending
to 4.571\%. Changing only payload size from 4 to 8 KiB yields 2,546.8 MiB/s
(+44.1\%) and 11,100 ms p99. Smaller qualified gains occur with 80 ms linger
(+4.8\%, 185 ms p99) and 60 partitions (+2.6\%, 188 ms p99). These single-run
contrasts identify follow-up priorities, without isolating causal improvements.

\subsection{Repeated validation and broker profiles}

\begin{table*}[!t]
\caption{Final Kafka validation under frozen B0.}
\label{tab:kafka-final}
\centering
\small
\setlength{\tabcolsep}{5pt}
\begin{tabular*}{\linewidth}{@{\extracolsep{\fill}}l c r r r r r@{}}
\toprule
Configuration & Q/5 & Median MiB/s & IQR & p99 ms & Pending \% & Flush s \\
\midrule
\texttt{cfg\_007} & 5/5 & 2,927.1 & 4.8 & 2,560 & 1.237 & 0.480 \\
\texttt{cfg\_062} & 5/5 & 2,842.1 & 99.4 & 6,510 & 1.870 & 1.449 \\
\texttt{cfg\_049} & 5/5 & 2,795.5 & 56.5 & 6,340 & 3.375 & 1.897 \\
\texttt{cfg\_096} & 5/5 & 2,622.5 & 27.2 & 270 & 0.170 & 0.076 \\
\texttt{cfg\_001} & 5/5 & 1,745.5 & 17.2 & 224 & 0.307 & 0.100 \\
\texttt{cfg\_094} & 5/5 & 1,002.2 & 3.0 & 756 & 1.146 & 0.391 \\
\texttt{cfg\_113} & 4/5 & 3,009.2 & 87.8 & 6,760 & 4.085 & 1.800 \\
\texttt{cfg\_107} & 4/5 & 3,001.7 & 151.7 & 7,620 & 3.982 & 2.086 \\
\texttt{cfg\_061} & 3/5 & 2,946.3 & 82.2 & 7,950 & 4.768 & 2.575 \\
\texttt{cfg\_101} & 0/5 & 2,868.3 & 171.3 & 11,400 & 10.454 & 5.450 \\
\bottomrule
\end{tabular*}

\par\smallskip
\begin{minipage}{\linewidth}\footnotesize
Q/5 counts qualified observations;
all five observations per configuration are eligible. Rate medians and IQRs refer to
the balanced endpoint rate in MiB/s;
p99, producer pending-delivery ratio, and flush are medians across observations. Median
failed-send percentages are zero throughout. Qualification is assessed per
observation, rather than from the medians. Endpoint minima and classifications
were recomputed from retained rows; p99 values come from retained per-run summaries.
Displayed decimals do not imply measurement accuracy.
\end{minipage}
\end{table*}

\begin{figure*}[!t]
\centering
\begin{tikzpicture}[font=\rmfamily\footnotesize]
\draw[mqblue,thick] (5.0,4.35) -- (5.5,4.35)
 node[right,text=black] {V1};
\fill[mqblue] (5.25,4.35) circle (2pt);
\draw[mqteal,thick] (7.0,4.35) -- (7.5,4.35)
 node[right,text=black] {V2};
\fill[mqteal] (7.18,4.28) rectangle (7.32,4.42);
\begin{scope}[x=1.15cm,y=0.30cm]
\fill[mqgrid!30] (3.5,0) rectangle (5.5,10);
\foreach \y in {0,2,4,6,8,10} {
 \draw[mqgrid] (0.5,\y) -- (5.5,\y);
 \node[left] at (0.45,\y) {\y};
}
\draw (0.5,10.3) -- (0.5,0) -- (5.5,0);
\draw[densely dashed,mqgray] (3.5,0) -- (3.5,10.3);
\foreach \x in {1,...,5} \node[below] at (\x,0) {\x};
\node[rotate=90] at (-0.18,5) {Qualified cases (of 10)};
\node at (3,-1.6) {Validation block};
\node at (2,10.8) {Batch 1};
\node at (4.5,10.8) {Batch 2};
\node at (3,12.2) {(a) Qualification by block};
% Verified from the ten case rows per block.
\draw[mqblue,thick] plot[mark=*,mark size=2pt] coordinates
 {(1,3) (2,3) (3,3) (4,9) (5,9)};
\draw[mqteal,thick] plot[mark=square*,mark size=2pt] coordinates
 {(1,9) (2,9) (3,8) (4,7) (5,8)};
\end{scope}
\begin{scope}[xshift=8cm,x=1.15cm,y=0.1875cm]
\fill[mqgrid!30] (3.5,0) rectangle (5.5,16);
\foreach \y in {0,5,10,15} {
 \draw[mqgrid] (0.5,\y) -- (5.5,\y);
 \node[left] at (0.45,\y) {\y};
}
\draw (0.5,16.5) -- (0.5,0) -- (5.5,0);
\draw[densely dashed,mqgray] (3.5,0) -- (3.5,16.5);
\draw[dashed,mqorange,thick] (0.5,5) -- (5.5,5);
\node[above,text=mqorange] at (2,5) {5\% policy limit};
\foreach \x in {1,...,5} \node[below] at (\x,0) {\x};
\node[rotate=90] at (-0.18,8) {Producer pending deliveries (\%)};
\node at (3,-2.56) {Validation block};
\node at (2,17.28) {Batch 1};
\node at (4.5,17.28) {Batch 2};
\node at (3,19.52) {(b) \texttt{cfg\_007} in each block};
% Verified from the cfg_007 pending-delivery counts and send attempts.
\draw[mqblue,thick] plot[mark=*,mark size=2pt] coordinates
 {(1,14.695) (2,14.567) (3,12.642) (4,3.815) (5,1.360)};
\draw[mqteal,thick] plot[mark=square*,mark size=2pt] coordinates
 {(1,1.469) (2,1.359) (3,1.237) (4,1.042) (5,1.072)};
\end{scope}
\end{tikzpicture}
\caption{Validation by block, recomputed from the retained case rows.
Each stage uses two batch allocations, separated by the vertical dashed line;
the V1 and V2 batches are distinct.  Panel (a) counts qualified workloads out
of ten; panel (b) shows one observation of \texttt{cfg\_007} per block.
The V1 change aligns with its allocation boundary, but this does not establish
an allocation effect.}
\label{fig:validation-blocks}
\end{figure*}
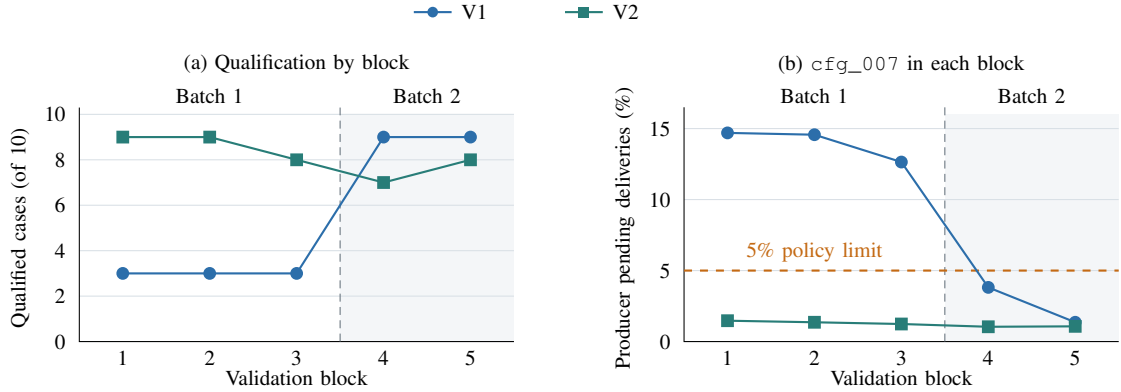

An instrumentation pilot compares sampling disabled with every-tenth sampling
in three pairs on \texttt{cfg\_074} (16 producer ranks, 64 consumer ranks,
2 KiB records). Median paired balanced-rate loss is 0.41\%, below the 3\%
pilot gate. This measures incremental latency instrumentation for one workload,
not total benchmark overhead or overhead at the highest rates.
Sampling-disabled cases use the pilot's separate validity rule.

V1's 50 validation observations are eligible, but only 27 qualify. Prioritizing
qualified count selects the baseline, with 5/5 qualified observations, median
1,746.4 MiB/s, and IQR 1.5 MiB/s. The screening byte-rate leader
\texttt{cfg\_107} and producer-rank candidate \texttt{cfg\_007} each qualify
in only 2/5 observations. A promising screen does not suffice for recommendation.

Profile screening compares B0--B5: network/I/O threads and heap GiB are
(8/16, 8), (16/32, 16), (32/64, 32), (48/96, 32), (48/96, 64), and
(48/96, 64), respectively. All use G1 GC. B4 and B5 share thread and heap settings;
B5 differs only in log-segment size ($2^{31}-1$ bytes versus 1 GiB). Each profile runs once on four
maximum-load anchors (\texttt{cfg\_107}, \texttt{cfg\_061}, \texttt{cfg\_094},
\texttt{cfg\_101}) and one fixed-rate latency anchor. B0, B3, and B5 receive
two additional observations per anchor. Decisions use three-run medians,
including eligible overdriven cases. The target offered rate for the fixed-load
latency anchor is 285,371 records/s, 80\% of the three-run calibration median. Its baseline client settings
change to 32 KiB batch, 1 ms linger, 1-byte fetch minimum, and 1 ms fetch wait.

Acceptance requires a geometric mean of matching B0-normalized balanced-rate
medians of at least 1.03 across the four maximum-load anchors, at most a 10\%
fixed-load p99 increase, no qualified-count reduction on \texttt{cfg\_107} and
\texttt{cfg\_061}, and no correctness or failed-send regression. Profile
identity/hash, required diagnostic samples, and tmpfs occupancy at most 75\%
must also pass. Final sampled request and response queues within the measurement
window must be empty; this is not a post-drain check.
The \texttt{iperf3} eligibility gate requires completed records with finite
positive values, without a stability or minimum-capacity requirement. Missing
diagnostics can thus exclude an otherwise valid application measurement.
B3 and B5 rate ratios are 1.0161 and 1.0157, below 1.03; fixed-load p99 ratios
are about 0.596 and 0.597. B0 is retained. The latency anchor is excluded from
the throughput geometric mean.

Final V2 validation again uses ten workloads and five blocks under B0: all
50 observations are eligible and 41 qualify. Table~\ref{tab:kafka-final} reports
the implemented ranking. \texttt{cfg\_007}, which changes only baseline producer
ranks from 40 to 64, qualifies 5/5 with median 2,927.1 MiB/s (749,340 records/s),
IQR 4.8 MiB/s, 1.237\% pending, and 0.480 s flush, with no failed sends.
The faster medians of \texttt{cfg\_113} and \texttt{cfg\_107} rank lower because
each has an overdriven observation. \texttt{cfg\_101} qualifies 0/5.

Reclassifying the fixed V2 shortlist at 2\%, 5\%, and 10\% pending limits
yields 25, 41, and 47 qualified observations; leaders are \texttt{cfg\_007},
\texttt{cfg\_007}, and \texttt{cfg\_113}. Other gates remain fixed; shortlists
are not regenerated under the alternate policies. Substituting $R_p$ for the
balanced score preserves the complete V2 order. Using $R_c$ only swaps
\texttt{cfg\_107} and \texttt{cfg\_113}, both 4/5-qualified; \texttt{cfg\_007}
remains first, with median $R_c=2{,}937.1$ MiB/s. Both endpoint alternatives
also preserve the V1 order.

The selected workload's median p99 is 2,560 ms versus the baseline's 224 ms.
The 5/5-qualified \texttt{cfg\_096} offers 2,622.5 MiB/s at 270 ms p99, but
its 16 KiB records reduce record rate to 167,838 records/s. For record rate,
\texttt{cfg\_094} reaches 1,026,286 records/s with 1 KiB payloads and 5/5
qualification. Lower latency requires an explicit selection constraint.

V1 and V2 retain the same B0, so their different leaders cannot be attributed
to broker-profile tuning. Using only qualified observations for the secondary
rate median preserves both leaders and the V2 order. Fig.~\ref{fig:validation-blocks}
shows substantial within-stage variation: V1 block qualification counts are
3, 3, 3, 9, 9, while V2 counts are 9, 9, 8, 7, 8. In V1, \texttt{cfg\_007}
pending deliveries fall from 12.642--14.695\% in the first batch to 3.815\%
and 1.360\% in the second. This aligns with an allocation boundary but does
not establish its cause. Five blocks are not five independent allocation-level
repetitions, and V2 selection does not establish cross-campaign stability.

\subsection{Resources and limits of bottleneck attribution}
All 50 V2 runs have the measurements required by the resource checks, including
29--30 process samples between the first producer measurement start and last
producer send-loop stop. JMX byte and handler-idle rates use one-minute windows;
GC uses 30-second rates. Selecting samples inside the application window does
not remove that smoothing. Table~\ref{tab:kafka-resources} shows six workloads.

\begin{table*}[!t]
\caption{Broker resources in final validation under B0.}
\label{tab:kafka-resources}
\centering
\small
\setlength{\tabcolsep}{5pt}
\begin{tabular*}{\linewidth}{@{\extracolsep{\fill}}lrrrrrr@{}}
\toprule
Config & CPU & Kafka in / out & Handler idle & Heap & GC & tmpfs max. \\
 & (\%) & (MiB/s) & (\%) & (decimal GB) & (ms/s) & (\%) \\
\midrule
\texttt{cfg\_001} & 5.14 & 1,797 / 1,772 & 93.4 & 2.78 & 3.31 & 32.1 \\
\texttt{cfg\_007} & 4.21 & 2,959 / 2,946 & 95.3 & 2.80 & 3.71 & 53.0 \\
\texttt{cfg\_107} & 3.64 & 3,164 / 3,180 & 95.2 & 2.74 & 3.64 & 56.1 \\
\texttt{cfg\_101} & 3.80 & 3,139 / 3,132 & 95.2 & 2.63 & 3.51 & 56.2 \\
\texttt{cfg\_096} & 4.39 & 2,619 / 2,612 & 94.2 & 2.67 & 3.69 & 47.1 \\
\texttt{cfg\_094} & 5.17 & 1,034 / 1,037 & 93.3 & 2.73 & 3.32 & 19.3 \\
\bottomrule
\end{tabular*}

\par\smallskip
\begin{minipage}{\linewidth}\footnotesize
Entries are medians
across five runs of each run's monitoring-window mean, except tmpfs, which
is the maximum over all five runs. CPU is broker-process utilization divided
by the logical CPUs allowed to that process. Kafka byte rates are JMX rates converted to MiB/s;
heap uses decimal GB. GC is the mean exported collector time rate.
The baseline and \texttt{cfg\_007}, \texttt{cfg\_096}, and \texttt{cfg\_094}
qualified in 5/5 observations, \texttt{cfg\_107} in 4/5, and \texttt{cfg\_101} in 0/5.
\end{minipage}
\end{table*}

Every V2 run has zero mean and zero 95th-percentile sampled request/response
queue sizes, with mean handler idle of 92.3--96.0\%. Sampling can miss bursts and
smoothing prevents exact window alignment. Normalized broker CPU can hide a
busy thread; client CPU and network-processor idle are unavailable. The
11.40 s application p99 of overdriven \texttt{cfg\_101} coexists with about
95\% handler idle. Averaged per-request-type JMX p99 values cannot be subtracted
from application p99 to locate delays. No unique bottleneck is isolated.

Mean heap use is 2.43--3.14 decimal GB across runs; the largest per-run 95th
percentile is 5.26 GB, below 8 GiB. Exported GC means of 3.22--4.04 ms/s are
neither total pause fractions nor maximum pauses. Maximum tmpfs occupancy is
56.46\%. These samples show no heap or log-space exhaustion but cannot exclude
transient pressure; RAM-backed logs preclude persistent-disk evaluation.

For \texttt{cfg\_007}/\texttt{cfg\_101}, broker \texttt{ib0} receive means are
3,057/3,272 MiB/s, including all interface traffic. Their JMX rates in
Table~\ref{tab:kafka-resources} and application rates in Table~\ref{tab:kafka-final}
have different byte scopes, smoothing, and endpoint windows; differences do
not establish packet loss.

The preflight TCP probes are unvalidated diagnostic metadata, not evidence of
data-path capacity. Producer-to-broker/broker-to-consumer rates are
10.134/9.968 Gbit/s for blocks 1--3 and 10.492/10.597 Gbit/s for blocks 4--5:
two allocation-level pairs, not 50 independent measurements. These values
conflict with roughly 3,000 MiB/s (25 Gbit/s) Kafka traffic. Missing raw probes
and routes prevent resolving the discrepancy. The positive-value gate does
not validate the path; these probes cannot support link-utilization or
bottleneck claims.

\section{Limitations and Evidence}
\label{sec:limitations}
The single-broker adapter is both an implementation and evaluation limitation.
Four-node role execution provides no evidence of broker-count scaling,
inter-broker replication, or partition balancing across brokers. RAM-backed
logs, replication factor one, short windows, and absence of failure-under-load
experiments limit generalization. Deterministic sampling can bias latency, while metadata and aggregate-count
checks cannot establish full payload integrity or globally unique delivery.
Qualification is a declared policy, not a capacity certificate.

Screening contrasts are single observations; each validation stage spans only
two allocations. Reusing block orders in V2 permits recurring temporal effects.
The selected V2 leader has no independent held-out confirmation. A further
campaign should replicate across new allocations with new orders, longer
windows, and complete runtime and network records. Randomized sampling and
overhead pilots at selected, low-latency, and high-rate workloads are also
needed to assess latency bias and instrumentation cost beyond the single pilot.

All 220 selected screening and validation rows were audited without mismatches
in recomputable quantities: endpoint minima, record-rate conversions, delivery
ratios, classifications from retained flags/counters, summaries, and block
counts. Retained per-run latency and resource summaries are not independently
reconstructed raw measurements. The campaign commit, exact Kafka/client
versions, full hardware inventory, raw per-rank times, histograms, clock
exchanges, and original network probes are unavailable. Hashes for 15 workflow
and analysis files match inspected source, without identifying the complete
campaign source revision. Current tests establish current implementation
behaviour, not campaign-time code identity.

\noindent\textbf{Code and data availability.}
\par\noindent\url{https://github.com/skysepehr/hpc-mqbench}
\par\noindent The repository provides benchmark code, retained Kafka case rows,
manifests, profile summaries, measurement contracts, and verification scripts.
It is not a complete raw campaign archive.

\section{Conclusion}
The benchmark connects Slurm experiment state to messaging validity, delivery
qualification, and configuration ranking. Its four-node single-broker evaluation
shows why the highest byte rate alone is insufficient. The V2-selected workload
qualifies in five observations at a median 2,927 MiB/s and 2.56 s p99, but qualifies
only twice in V1 under the same broker profile. This supports an auditable
stage-specific decision, not a stable Kafka recommendation or sustained-capacity
claim. Multi-broker deployment remains future work, requiring cluster membership,
partition/replica placement, per-broker health and aggregated monitoring, followed
by experiments varying broker count with documented client and replication settings.

\section*{Acknowledgment}
The authors gratefully acknowledge computing time on Emmy at
NHR-Nord@G\"ottingen, jointly supported by the Federal Ministry of Research,
Technology and Space and participating state governments through the
\href{https://www.nhr-verein.de/nhr-alliance/nhr-funding/}{National High-Performance Computing (NHR) joint funding program}.

OpenAI Codex assisted in drafting and revising text throughout the manuscript
and in preparing LaTeX and figure layouts. Reported numerical results derive
from retained benchmark evidence.

\IEEEtriggeratref{15}
\bibliographystyle{IEEEtran}
\bibliography{references}
\end{document}